\documentstyle[epsfig]{mn}
\newcommand{\reference}{\bibitem}
\title[]{On the Optical Light Curves of Afterglows from 
         Jetted Gamma-ray Burst Ejecta: Effects of Parameters} 

\author[]{Y.F. Huang$^{1,2}$, Z.G. Dai$^{1,3}$ and T. Lu$^{1,3}$ 
\thanks{E-mail: tlu@nju.edu.cn} \\
$^1${\sl Department of Astronomy, Nanjing University, Nanjing 210093, 
         P. R. China} \\
$^2${\sl Astronomical and Astrophysical Center of East China, 
         Nanjing University, Nanjing 210093, P. R. China; hyf@nju.edu.cn} \\
$^3${\sl LCRHEA, Institute for High-Energy Physics, Chinese Academy of 
         Sciences, Beijing 100039, P. R. China}} 
\date{MNRAS in press }
\pubyear{1999}

\begin{document}
\voffset=-0.5 in

\maketitle
\begin{abstract}
Due to some refinements in the dynamics, we can follow the overall 
evolution of a realistic jet numerically till its bulk velocity 
being as small as $\beta c \sim 10^{-3} c$. We find no obvious break 
in the optical light curve during the {\em relativistic phase 
itself}. However, an obvious break does exist at the transition 
from the relativistic phase to the non-relativistic phase, which 
typically occurs at time $t \sim 10^6$ --- $10^{6.5}$ s 
(i.e., 10 --- 30 d). The break is affected by many parameters, 
such as the electron energy fraction $\xi_{\rm e}$, the magnetic 
energy fraction $\xi_{\rm B}^2$, the initial half opening angle 
$\theta_0$, and the medium number density $n$. Increase of any of 
them to a large enough value will make the break disappear. Although
the break itself is parameter-dependent, afterglows from jetted GRB 
remnants are uniformly characterized by a quick decay during the 
non-relativistic phase, with power law timing index $\alpha \geq 2.1$. 
This is quite different from that of isotropic fireballs, and may be 
of fundamental importance for determining the degree of beaming in 
$\gamma$-ray bursts observationally. 
\end{abstract}

\section {Introduction}

The discovery of afterglows from gamma-ray bursts (GRBs) has opened up 
a new era in the field. Till the end of August 1999, X-ray, optical, 
and radio afterglows have been observed from about 16, 11, and 5 GRBs 
respectively (Costa et al. 1997; Bloom et al. 1998; Groot et al. 1998; 
Kulkarni et al. 1998, 1999; Harrison et al. 1999; Stanek et al. 1999; 
Fruchter et al. 1999; Galama et al. 1999a). The so called fireball 
model (Goodman 1986; Paczy\'{n}ski 1986; M\'{e}sz\'{a}ros \& Rees 1992; 
Rees \& M\'{e}sz\'{a}ros 1992, 1994; Katz 1994; Sari, Narayan \& Piran 
1996) is strongly favored, which is found successful at explaining the 
major features  of GRB afterglows (M\'{e}sz\'{a}ros \& Rees 1997; 
Vietri 1997; Tavani 1997; Waxman 1997a; Wijers, Rees \& M\'{e}sz\'{a}ros 
1997; Sari 1997a; Sari, Piran \& Narayan 1998; Huang et al. 1998a, b, 
1999a, b; Dai \& Lu 1998a, b, c; M\'{e}sz\'{a}ros, Rees \& Wijers 1998; 
Wijers \& Galama 1999).  However, we are still far from resolving the 
puzzle of GRBs, because their ``inner engines'' are well hidden from 
direct afterglow observations.

Most GRBs localized by BeppoSAX have indicated isotropic energies of 
$10^{51}$ --- $10^{52}$ ergs, well within the energy output from 
solar-mass compact stellar objects. However, 
GRB 971214, 980703, 990123, and 990510 have implied isotropic 
gamma-ray releases of $3.0 \times 10^{53}$ ergs (0.17 $M_{\odot} c^2$,
Kulkarni et al. 1998), $1.0 \times 10^{53}$ ergs (0.06 $M_{\odot} c^2$, 
Bloom et al. 1998), $3.4 \times 10^{54}$ ergs (1.9 $M_{\odot} c^2$, 
Kulkarni et al. 1999; Andersen et al. 1999), 
and $2.9 \times 10^{53}$ ergs 
(0.16 $M_{\odot} c^2$, Harrison et al. 1999) respectively. Moreover, 
if really located at a redshift of $z \geq 5$ as suggested by 
Reichart et al. (1998), GRB 980329 would imply an isotropic gamma-ray 
energy of $5 \times 10^{54}$ ergs (2.79 $M_{\odot} c^2$). Such enormous 
energetics has forced some theorists to deduce that GRB radiation must 
be highly collimated in these cases, with half opening 
angle $\theta \leq 0.2$, so that the intrinsic gamma-ray energy could be 
reduced by a factor of $10^2$ --- $10^3$, and could still come from 
compact stellar objects (Pugliese, Falcke \& Biermann 1999). 
Obviously, whether GRBs are beamed or not is of 
fundamental importance to our understanding of their nature. For theorists, 
the degree of beaming can place severe constraints on GRB models.

How can we tell a jet from an isotropic fireball? 
Direct clues may come from afterglow light curves. As argued by 
Panaitescu \& M\'{e}sz\'{a}ros (1999) and 
Kulkarni et al. (1999), when the Lorentz factor of the ejecta 
drops to $\gamma < 1/ \theta$, the edge of the jet becomes visible.
Thus the light curve will steepen by $t^{-3/4}$, where $t$ is 
the observed time. This is called the edge effect (M\'{e}sz\'{a}ros 
\& Rees 1999). Another effect is called the lateral expansion effect.  
Rhoads (1997, 1999a, b) has shown that the lateral expansion 
(at sound speed) of a relativistic jet ($\gamma \geq 2$) will 
cause the blastwave to decelerate more quickly, leading to a sharp 
break in the afterglow light curve. The breaking point is again 
determined by $\gamma \sim 1/\theta$. The power law decay indices 
of afterglows from GRB 980326 and 980519 are anomalously 
large, $\alpha \sim 2.0$ (Groot et al. 1998; Owens et al. 1998; 
Halpern et al. 1999), and optical light curves of GRB 990123 and 
990510 even show obvious steepening at  
$t \geq 1$ --- 2 d (Kulkarni et al. 1999; Harrison et al. 1999; 
Castro-Tirado et al. 1999). Recently GRB 970228 was also reported to 
have a large index of $\alpha \sim  1.73$ (Galama et al. 1999b).
These phenomena have been widely regarded as evidence for 
relativistic jets (Sari, Piran \& Halpern 1999).  

However, numerical studies of some other authors (Panaitescu \& 
M\'{e}sz\'{a}ros 1998; Moderski, Sikora \& Bulik 1999) have shown 
that due to the increased swept-up matter and the time delay of the 
large angle emission, the sideway expansion of the jet does not lead 
to an obvious dimming of the afterglow. Thus there are two opposite 
conclusions about the jet effect: the analytical solution predicts a 
sharp break, while the numerical calculation shows no such sharp 
breaks. The condition is quite confusing. We need to clarify this 
question urgently.
 
In a recent paper (Huang et al. 1999c), we have developed 
a refined model to describe the evolution of jetted GRB remnants. 
Due to some crucial refinements in the dynamics, we can follow the 
overall evolution of a realistic jet till its expanding velocity is 
as small as $\sim 10^{-3} c$. Many new results were obtained in that 
paper, e.g., (i) We found no obvious break in the optical 
light curve during {\em the relativistic phase itself}, i.e. the 
time determined by $\gamma \sim 1/\theta$ is not a breaking point.
But in some cases, obvious breaks does appear at the 
relativistic-Newtonian transition point. 
(ii) Generally speaking, the Newtonian phase of jet evolution is 
characterized by a sharp decay of optical afterglows, with the power
law timing index $\alpha \geq 1.8$ --- $2.1$\,. The most interesting 
finding may be that whether the relativistic-Newtonian break appears 
or not depends on $\xi_{\rm e}$, the parameter characterizing 
the energy equipartition between electrons and protons. This has 
given strong hints on the solution to the confusing problem 
mentioned just above: whether an obvious break appears or not may 
depend on parameters.

In this paper, we go further to investigate what impact will other 
parameters have on the optical light curves, based on the model developed 
by Huang et al. (1999c). The organization of the paper is as follows.  
For completeness,  
the model is briefly described in Section 2. In Section 3 we investigate 
various parameter effects and present our detailed numerical results, 
mainly in the form of optical light curves. 
We find that the light curve break is really affected by many other 
parameters. Section 4 is our final conclusion, 
and Section 5 is a brief discussion.

\section {Model}

We use the model developed by Huang et al. (1999c). This model has the 
following advantages: (i) It is applicable to both radiative and adiabatic 
blastwaves, and appropriate for both ultra-relativistic and non-relativistic 
stages. The model even allows the radiative efficiency $\epsilon$ to 
evolve with time, so that it can trace the evolution of a realistic 
GRB remnant, which is believed to evolve from the highly radiative regime
to the adiabatic one (Dai, Huang \& Lu 1999). (ii) It takes the 
lateral expansion of the jet into account. The lateral speed is given 
by a reasonable expression. (iii) It also takes many other effects into 
account, for example, the cooling of electrons, and the equal arrival 
time surfaces. The model is very convenient for numerical studies. Here, 
for completeness,  
we describe the model briefly. For details please see 
Huang et al. (1999c).

\subsection{Dynamics}

Let $R$ be the radial coordinate in the burster frame; $t$ be the 
observer's time; $\gamma_0$ and $M_{\rm ej}$ be the initial Lorentz 
factor and ejecta mass and $\theta$ the half opening angle of the 
ejecta. The burst energy is $E_0 = \gamma_0 M_{\rm ej} c^2$. 

The evolution of radius ($R$), the swept-up mass ($m$), the half 
opening angle ($\theta$) and the Lorentz factor ($\gamma$) is described 
by (Huang et al. 1999c):
\begin{equation}
\label{drdt1}
\frac{d R}{d t} = \beta c \gamma (\gamma + \sqrt{\gamma^2 - 1}),
\end{equation}
\begin{equation}
\label{dmdr2}
\frac{d m}{d R} = 2 \pi R^2 (1 - \cos \theta) n m_{\rm p},
\end{equation}
\begin{equation}
\label{dthdt3}
\frac{d \theta}{d t} \equiv \frac{1}{R} \frac{d a}{d t}
		     = \frac{c_{\rm s} (\gamma + \sqrt{\gamma^2 - 1})}{R},
\end{equation}
\begin{equation}
\label{dgdm4}
\frac{d \gamma}{d m} = - \frac{\gamma^2 - 1}
       {M_{\rm ej} + \epsilon m + 2 ( 1 - \epsilon) \gamma m}, 
\end{equation}
where $\beta = \sqrt{\gamma^2-1}/\gamma$, $n$ is the number density of 
surrounding interstellar medium (ISM), $m_{\rm p}$ is the mass of proton, 
$a$ is the co-moving lateral radius of the ejecta (Rhoads 1999a; Moderski, 
Sikora \& Bulik 1999), $c_{\rm s}$ is the co-moving sound speed, 
and $\epsilon$ is the radiative efficiency.

A reasonable expression for $c_{\rm s}$ is 
\begin{equation}
\label{cs5}
c_{\rm s}^2 = \hat{\gamma} (\hat{\gamma} - 1) (\gamma - 1) 
	      \frac{1}{1 + \hat{\gamma}(\gamma - 1)} c^2 , 
\end{equation}
where $\hat{\gamma} \approx (4 \gamma + 1)/(3 \gamma)$ is the adiabatic 
index. In the ultra-relativistic limit 
($\gamma \gg 1, \hat{\gamma} \approx 4/3$), 
equation~(\ref{cs5}) gives $c_{\rm s}^2 = c^2/3$; and in the 
non-relativistic limit ($\gamma \sim 1, \hat{\gamma} \approx 5/3$), we 
simply get $c_{\rm s}^2 = 5 \beta^2 c^2/9$. 

As usual, we assume that the magnetic energy density in the co-moving 
frame is a fraction $\xi_{\rm B}^2$ of the total thermal energy density 
(Dai, Huang \& Lu 1999), $B'^2 / 8 \pi = \xi_{\rm B}^2  e'$, 
and that the shock accelerated electrons carry a fraction $\xi_{\rm e}$ of 
the proton energy. The minimum Lorentz factor of the 
random motion of electrons in the co-moving frame is 
$\gamma_{\rm e,min} = \xi_{\rm e} (\gamma - 1) 
		     m_{\rm p} (p - 2) / [m_{\rm e} (p - 1)] + 1$, 
where $p$ is the index characterizing the power law energy distribution 
of electrons, and $m_{\rm e}$ is the electron mass. 
Then the radiative efficiency of the ejecta is (Dai, Huang \& Lu 1999)
\begin{equation}
\label{eps6}
\epsilon=\xi_{\rm e} \frac{t^{\prime -1}_{\rm syn}}{t^{\prime -1}_{\rm syn}
	 +t^{\prime -1}_{\rm ex}} ,  
\end{equation}
where $t^\prime_{\rm ex}$ is the co-moving frame expansion time,
$t^\prime_{\rm ex}=R/(\gamma c)$, and $t^\prime_{\rm syn}$ is the 
synchrotron cooling time, $t^\prime_{\rm syn}=
6\pi m_{\rm e} c/(\sigma_{\rm T}B^{\prime 2}\gamma_{\rm e,min})$,
with $\sigma_{\rm T}$ the Thompson cross section. In this paper, we 
call the jet, whose radiative efficiency evolves according to 
equation~(\ref{eps6}), a ``realistic'' one (Dai, Huang \& Lu 1999).  

\subsection{Synchrotron Radiation}

In the co-moving frame, synchrotron radiation power at frequency $\nu '$ from 
electrons is given by (Rybicki \& Lightman 1979)
\begin{equation}
\label{pnue7}
P'(\nu ') = \frac{\sqrt{3} e^3 B'}{m_{\rm e} c^2} 
	    \int_{\gamma_{\rm e,min}}^{\gamma_{\rm e,max}} 
	    \left( \frac{dN_{\rm e}'}{d\gamma_{\rm e}} \right)
	    F\left(\frac{\nu '}{\nu_{\rm c}'} \right) d\gamma_{\rm e},
\end{equation}
where $e$ is electron charge, $\gamma_{\rm e,max} = 
10^8 (B'/1{\rm G})^{-1/2}$ is the maximum Lorentz factor of electrons, 
$dN_{\rm e}'/d\gamma_{\rm e}$ is electron distribution function, 
$\nu_{\rm c}' = 3 \gamma_{\rm e}^2 e B' / (4 \pi m_{\rm e} c)$, and 
\begin{equation}
\label{fx8}
F(x) = x \int_{x}^{+ \infty} K_{5/3}(k) dk,
\end{equation}
with $K_{5/3}(k)$ being the Bessel function. 

In the absence of radiation loss, the distribution of the shock accelerated 
electrons behind the blastwave is usually assumed to be a power law function 
of electron energy, $dN_{\rm e}' / d\gamma_{\rm e} \propto 
\gamma_{\rm e}^{-p}$. However, radiation loss may play an important role in 
the process. Sari, Piran \& Narayan (1998) have derived an equation for 
the critical electron Lorentz factor, $\gamma_{\rm c}$,
above which synchrotron radiation is significant:
$\gamma_{\rm c}=6 \pi m_{\rm e} c / (\sigma_{\rm T} \gamma B'^2 t)$.
In our model, the actual electron distribution is given according to 
the following cases (Dai, Huang \& Lu 1999; Huang et al. 1999c):
\begin{description}
\item (i) For $\gamma_{\rm c}\leq \gamma_{\rm e,min}$,
\begin{equation}
\label{dne9}
\frac{dN_{\rm e}'}{d\gamma_{\rm e}} \propto \gamma_{\rm e}^{-(p+1)}\,, \,\,\,\,
(\gamma_{\rm e,min}\leq\gamma_{\rm e}\leq \gamma_{\rm e,max});
\end{equation}

\item (ii) For $\gamma_{\rm e,min} < \gamma_{\rm c} \leq \gamma_{\rm e,max}$,
\begin{equation}
\label{dne10}
  \frac{dN_{\rm e}'}{d\gamma_{\rm e}} \propto \left \{
   \begin{array}{ll}
 \gamma_{\rm e}^{-p}\,, \,\,\,\, & (\gamma_{\rm e,min} \leq \gamma_{\rm e}
					      \leq \gamma_{\rm c}), \\
 \gamma_{\rm e}^{-(p+1)}\,, \,\,\,\, & (\gamma_{\rm c}<\gamma_{\rm e}
					      \leq \gamma_{\rm e,max});
   \end{array}
   \right. 
\end{equation}

\item (iii) For $\gamma_{\rm c} > \gamma_{\rm e,max}$, 
\begin{equation}
\label{dne11}
\frac{dN_{\rm e}'}{d\gamma_{\rm e}} \propto \gamma_{\rm e}^{-p}, \,\,\,\,
(\gamma_{\rm e,min}\leq\gamma_{\rm e}\leq\gamma_{\rm e,max}).
\end{equation}
\end{description}

Let $D_{\rm L}$ be the luminosity distance, $\Theta$ be the angle 
between the velocity of emitting material and the line of sight and 
define $\mu = \cos \Theta$. Then the observed flux density at 
frequency $\nu$  from this emitting point is 
\begin{equation}
\label{snu12}
S_{\nu} = \frac{1}{\gamma^4 (1 - \beta \mu)^3} \frac{1}{4 \pi D_{\rm L}^2}
          P'\left(\gamma(1 - \mu \beta) \nu \right).
\end{equation}
In order to calculate the total observed flux density, we should integrate 
over the equal arrival time surface (Waxman 1997b; Sari 1997b; 
Panaitescu \& M\'{e}sz\'{a}ros 1998) determined by 
\begin{equation}
\label{eqt28}
t = \int \frac{1 - \beta \mu}{\beta c} dR \equiv {\rm const},
\end{equation}
within the jet boundaries (Moderski, Sikora \& Bulik 1999;
Panaitescu \& M\'{e}sz\'{a}ros 1999). 

\begin{figure} \centering
\epsfig{file=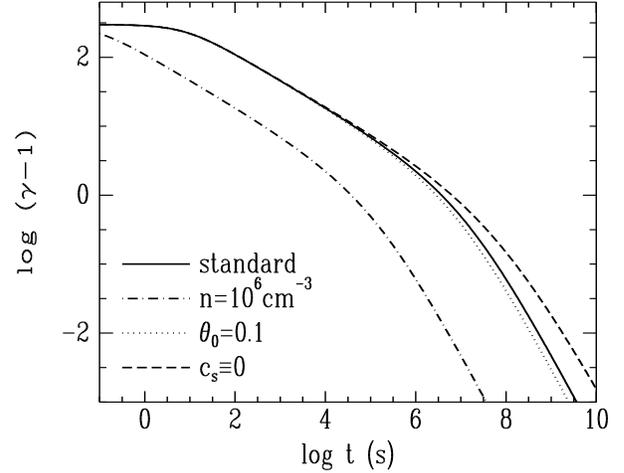, angle=-90, height=7cm, width=8.5cm,
bbllx=130pt, bblly=70pt, bburx=570pt, bbury=640pt}
\caption{The evolution of the Lorentz factor $\gamma$. The solid line 
corresponds to a jet with ``standard'' parameters. Other lines are drawn 
with only one parameter altered or one condition changed with respect to 
the solid line. The dashed line corresponds to a jet without lateral 
expansion, i.e., $c_{\rm s} \equiv 0$ cm/s; the dotted
line is for $\theta_0 = 0.1$; and the dash-dotted line is for $n = 10^6$
cm$^{-3}$. For the meaning of ``standard'', see Section 3 in the 
main text.}
\end{figure}

\begin{figure} \centering
\epsfig{file=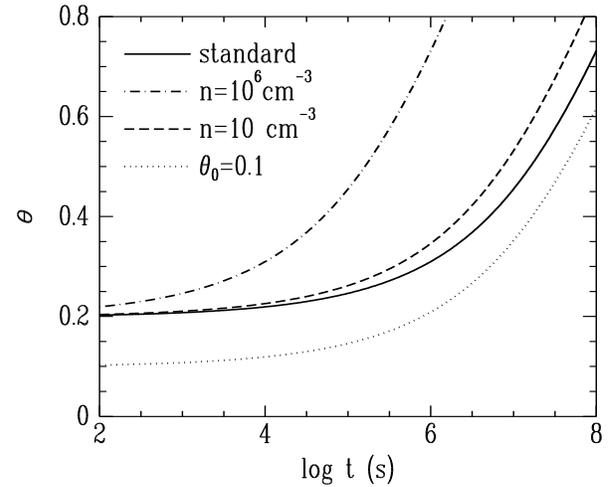, angle=-90, height=7cm, width=8.5cm,
bbllx=130pt, bblly=70pt, bburx=570pt, bbury=640pt}
\caption{The evolution of the half opening angle $\theta$. The solid line
corresponds to a jet with ``standard'' parameters. Other lines are drawn with
only one parameter altered.}
\end{figure}

\begin{figure} \centering
\epsfig{file=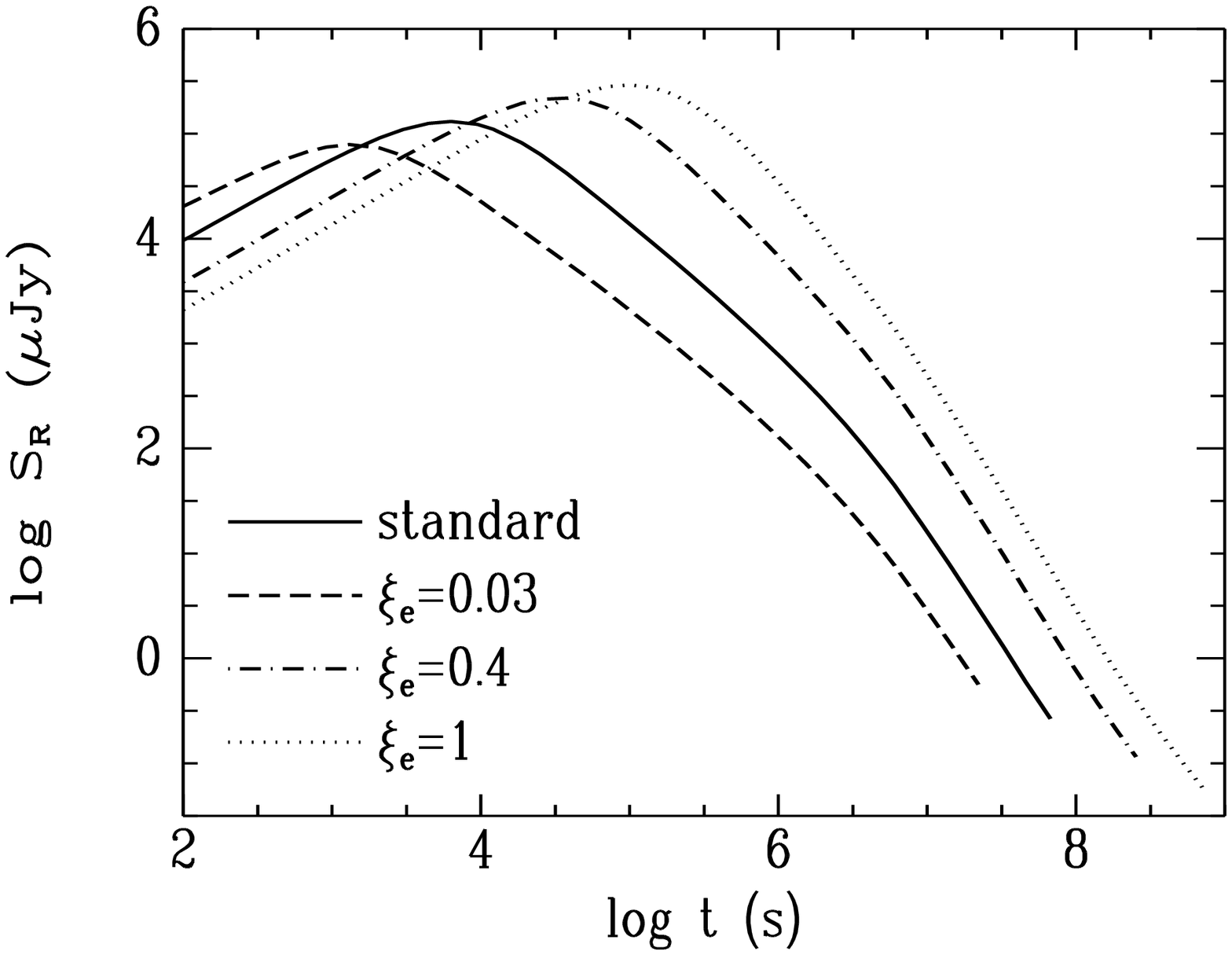, angle=-90, height=7cm, width=8.5cm,
bbllx=130pt, bblly=70pt, bburx=570pt, bbury=640pt}
\caption{The effect of the parameter $\xi_{\rm e}$ on the optical light 
curve. The solid line corresponds to a jet with ``standard'' parameters.
Other lines are drawn with only $\xi_{\rm e}$ altered.}
\end{figure}

\begin{figure} \centering
\epsfig{file=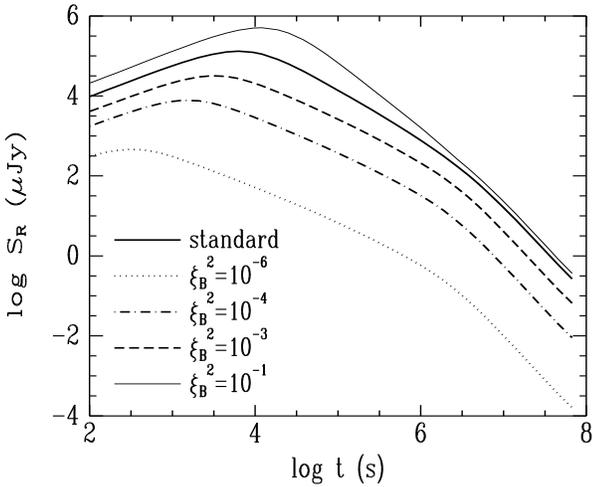, angle=-90, height=7cm, width=8.5cm,
bbllx=130pt, bblly=70pt, bburx=570pt, bbury=640pt}
\caption{The effect of the parameter $\xi_{\rm B}^2$ on the
optical light curve. The thick solid line corresponds to a jet with
``standard'' parameters. Other lines are drawn with only $\xi_{\rm B}^2$
altered.}
\end{figure}

\section{Numerical Results}

In this section, we follow the evolution of jetted GRB remnants 
numerically to see what effects will those parameters (such as 
$\xi_{\rm e}, \xi_{\rm B}^2, \theta_0, \theta_{\rm obs}, n, p,$
..., etc.) have on the optical light curves. 
For convenience, let us define the following initial values or parameters 
as a set of ``standard'' parameters: initial energy per solid 
angle $E_0 / \Omega_0 = 10^{54}$ ergs/$4 \pi$, 
$\gamma_0 = 300$ (i.e., initial ejecta mass per 
solid angle $M_{\rm ej}/\Omega_0 \approx 0.0019 M_{\odot}/ 4 \pi$), 
$n = 1$ cm$^{-3}$, $\xi_{\rm B}^2 = 0.01$, $p = 2.5$, 
$D_{\rm L} = 1.0 \times 10^6$ kpc, $\xi_{\rm e} =0.1$, $\theta_0 = 0.2, 
\theta_{\rm obs}=0$, where the observing angle $\theta_{\rm obs}$ is 
defined as the angle between the line of sight and the jet axis. 
For simplicity, we first assume that the expansion is completely 
adiabatic all the time (i.e. $\epsilon \equiv 0$, we call it an 
``ideal'' jet, distinguishing it from the ``realistic'' jet defined 
in Section 2.1).

Figure 1 shows the evolution of the Lorentz factor for some 
exemplary jets. In the ``standard'' case, the ultra-relativistic 
phase lasts only for $\sim 10^5$ s, and the non-relativistic phase 
begins at about $t \sim 10^{6.5}$ s. In short, the ejecta will cease 
to be highly relativistic at time $t \sim 10^5$ --- $10^6$ s. 
This again gives strong support to our previous argument that we should
be careful in discussing the fireball evolution under the simple 
assumption of ultra-relativistic limit (Huang et al. 1998a, b, 1999a, b). 
In the case of a dense ISM ($n = 10^6$ cm$^{-3}$, the dash-dotted line), 
the expansion will become non-relativistic as early as $t \sim 10^{4.5}$ s. 
Figure 2 is the evolution of the jet opening angle $\theta$. During 
the ultra-relativistic phase, $\theta$ increases only slightly. But 
at the Newtonian stage, the increase of $\theta$ is very quick.

\begin{figure} \centering
\epsfig{file=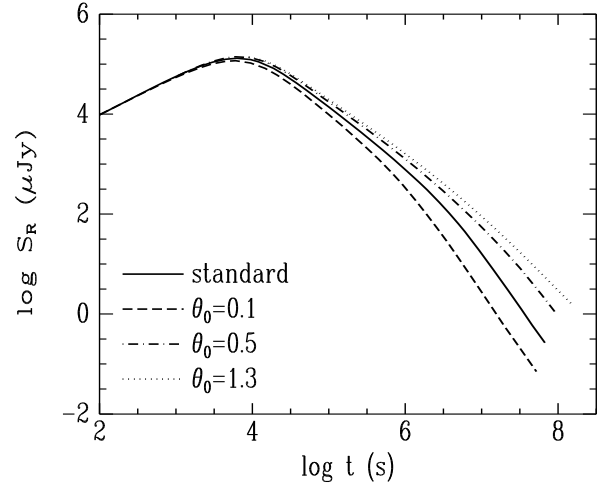, angle=-90, height=7cm, width=8.5cm,
bbllx=130pt, bblly=70pt, bburx=570pt, bbury=640pt}
\caption{The effect of the parameter $\theta_0$ on the optical
light curve. The solid line corresponds to a jet with ``standard'' parameters.
Other lines are drawn with only $\theta_0$ altered.}
\end{figure}

\begin{figure} \centering
\epsfig{file=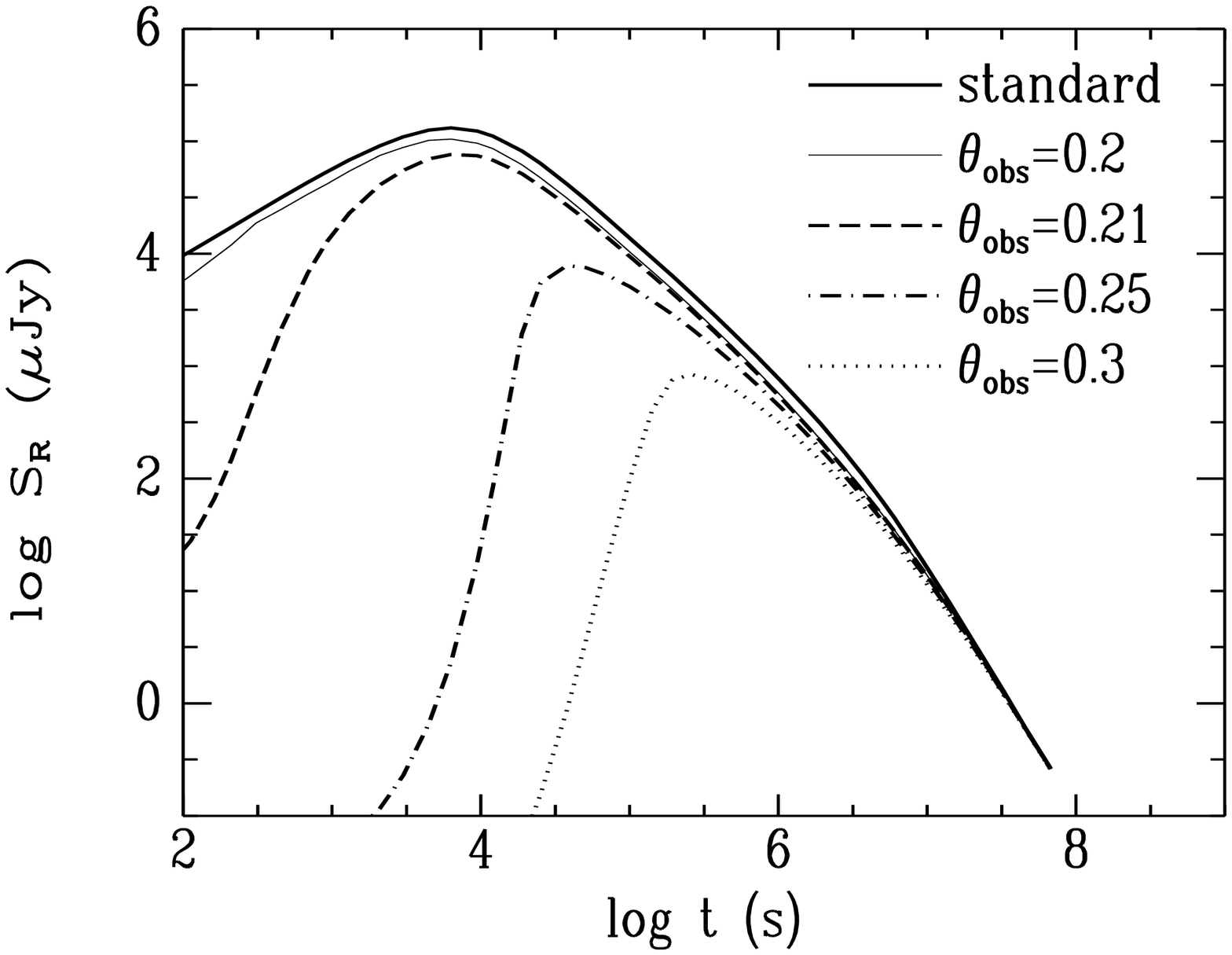, angle=-90, height=7cm, width=8.5cm,
bbllx=130pt, bblly=70pt, bburx=570pt, bbury=640pt}
\caption{The effect of the parameter $\theta_{\rm obs}$ on the optical
light curve. The solid line corresponds to a jet with ``standard'' parameters.
Other lines are drawn with only $\theta_{\rm obs}$ altered.}
\end{figure}

\begin{figure} \centering
\epsfig{file=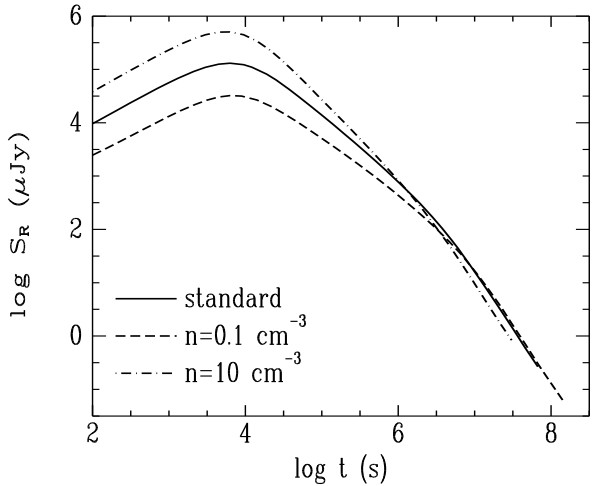, angle=-90, height=7cm, width=8.5cm,
bbllx=130pt, bblly=70pt, bburx=570pt, bbury=640pt}
\caption{The effect of the parameter $n$ on the optical
light curves. The solid line corresponds to a jet
with ``standard'' parameters, other lines are drawn 
with only $n$ altered.}
\end{figure}

The effect of $\xi_{\rm e}$ on the optical (R-band) light curves is 
illustrated in Figure 3, from which we can see clearly that: 
(i) In no case could we observe the theoretically predicted light 
curve steepening (with the break point determined by $\gamma \sim 
1/\theta$) during {\em the relativistic stage itself}, this is 
consistent with the result of Moderski, Sikora \& Bulik (1999). Note 
that in this figure the relativistic phase is restricted by 
$t \leq 10^6$ s. Huang et al. (1999c) have given a reasonable 
explanation to this phenomena: the edge and the lateral expansion 
effects begin to take effect when $\gamma \sim 1/\theta$ (in our 
calculation this occurs at $t \sim 10^{5.5}$ s, however the 
blastwave is already in its mildly relativistic phase at that 
moment and it will become non-relativistic soon after that (i.e., 
when $t \geq 10^{6.5}$ s, see Figure 1). So it is not surprising 
that we could not see any obvious breaks during the relativistic 
phase, they just do not have time to emerge. Another possible reason 
is: at these stages, since $\gamma$ is no longer much larger 
than 1, we should be careful in using the analytic approximations;
(ii) When $\xi_{\rm e}$ is small, an obvious break does appear in 
the light curve, but it is clearly due to the transition from 
the relativistic phase to the non-relativistic phase. The simulation 
by Moderski, Sikora \& Bulik (1999) does not show such breaks, 
because their model is not appropriate for non-relativistic expansion 
(Huang et al. 1999c); (iii) When $\xi_{\rm e}$ is large, the break 
disappears. This is not difficult to understand (Huang et al. 1999c). 
According to the analysis in the ultra-relativistic limit, the 
time that the light curve peaks scales as (Wijers \& Galama 1999; 
B\"{o}ttcher \& Dermer 1999; Chevalier \& Li 1999) 
\begin{equation}
\label{tm14}
t_{\rm m} \propto \left( \frac{p-2}{p-1} \right)^{4/3} 
                  \xi_{\rm e}^{4/3} (\xi_{\rm B}^2)^{1/3}.
\end{equation}
>From Figure 3 we can also see that with the increase of $\xi_{\rm e}$, 
$t_{\rm m}$ becomes larger and larger, consistent with 
equation~(\ref{tm14}). In the case of $\xi_{\rm e} = 1.0$, $t_{\rm m}$ 
is as large as $\sim 10^5$ s. Note that the expansion has already 
ceased to be ultra-relativistic at that moment. Then we can not see 
the initial power law decay (i.e., with $\alpha \sim 1.1$) in the 
relativistic phase, and so it is not strange that the break due to the
relativistic-Newtonian transition does not appear (Huang et al. 1999c); 
(iv) In all cases, the light curves during the non-relativistic phase 
are characterized by quick decays, with $\alpha \geq 2.1$. 
This is quite different from isotropic fireballs, whose light curves 
steepen only slightly after entering the non-relativistic phase 
(Wijers, Rees \& M\'{e}sz\'{a}ros 1997; Huang, Dai \& Lu 1998a). 
We thus suggest that 
the most obvious characteristic of jet effects is a sharp decay of 
afterglows at late stages (with $\alpha \geq 2$). This will be further
proved by other figures followed. 

\begin{figure} \centering
\epsfig{file=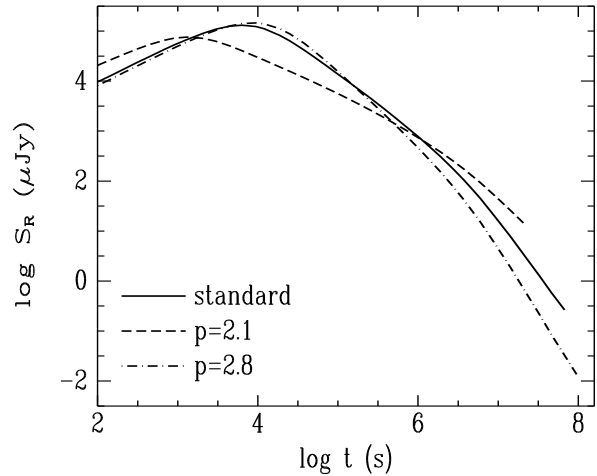, angle=-90, height=7cm, width=8.5cm,
bbllx=130pt, bblly=70pt, bburx=570pt, bbury=640pt}
\caption{The effect of the parameter $p$ on the optical
light curves. The solid line corresponds to a jet with ``standard''
parameters, other lines are drawn with only $p$ altered.}
\end{figure}

Figure 4 illustrates the effect of $\xi_{\rm B}^2$ on the optical 
light curves. Again no break appears during {\em the relativistic 
phase itself}. Interestingly but not surprisingly, we see that 
$\xi_{\rm B}^2$  has an effect similar to $\xi_{\rm e}$: for small 
values of $\xi_{\rm B}^2$, there are obvious breaks at the transition 
from relativistic stage to non-relativistic stage (i.e., at 
$t \sim 10^{6.5}$ s); but for large $\xi_{\rm B}^2$ values, the break 
disappears, we could only observe a single steep line with $\alpha \geq 2.1$. 
The reason is also similar to that of $\xi_{\rm e}$. With the increase 
of $\xi_{\rm B}^2$, $t_{\rm m}$ becomes larger and larger, as can be 
clearly seen from equation~(\ref{tm14}). When $t_{\rm m}$ is large 
enough (i.e., enters the mildly relativistic zone), the initial power 
law decaying segment (with $\alpha \sim 1.1$) in the relativistic 
phase will be hidden completely. Then we can only see the quick decay 
in the Newtonian phase. The figure also shows that with the decrease 
of $\xi_{\rm B}^2$, the flux density decreases substantially. 

Figure 5 illustrates the effect of $\theta_0$ on the light curves. For 
small $\theta_0$ values, the breaks due to the relativistic-Newtonian 
transition are very obvious, consistent with a recent analytic treatment 
by Wei \& Lu (1999). For large $\theta_0$ values, the breaks are not 
striking. The dotted line is drawn with $\theta_0 = 1.3$, it in fact 
can be regarded as an isotropic fireball. The most notable difference 
between the solid line and the dotted line is their slope disparity in 
the non-relativistic phase. This is an important difference 
between jet and isotropic fireball, and may be useful in determining the
degree of beaming.

The effect of $\theta_{\rm obs}$ on the light curves is shown in 
Figure 6. The thick solid line is drawn with $\theta_{\rm obs} = 0$. 
The light curve with $\theta_{\rm obs} = \theta_0$ 
deviates the thick solid line only slightly. But for $\theta_{\rm obs}
> \theta_0$, the observed flux decreases severely.
For $\theta_{\rm obs} = 0.3$, the observed peak flux density is 
dimmer than that of $\theta_{\rm obs} = 0$ by 5  mag. But fortunately, 
their late time behavior is very similar. So it is still possible 
for us to resort to systematic deep optical surveys, which are expected 
to find many faint afterglows not associated with any know GRBs if 
GRBs are highly collimated. 

\begin{figure} \centering
\epsfig{file=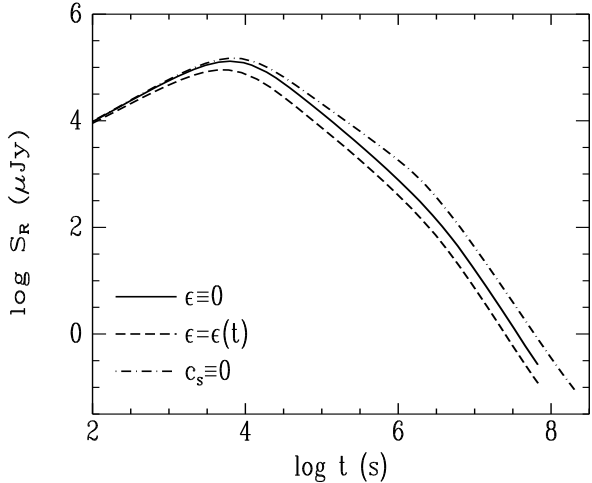, angle=-90, height=7cm, width=8.5cm,
bbllx=130pt, bblly=70pt, bburx=570pt, bbury=640pt}
\caption{Optical light curves under different assumptions.
All the lines are drawn with ``standard'' parameters.
The solid line corresponds to an adiabatic jet, while the
dash-dotted line corresponds to a jet without lateral expansion, 
and the dashed line corresponds to a jet that evolves from 
highly radiative regime to adiabatic regime.}
\end{figure}

Figure 7 illustrates the effect of ISM number density $n$. According to 
analysis under ultra-relativistic assumption, $t_{\rm m}$ is not related 
to $n$ ($t_{\rm m} \propto n^0$), and the peak flux density scales as: 
$S_{\rm R,m} \propto n^{1/2}$ (Wijers \& Galama 1999; B\"{o}ttcher \& 
Dermer 1999). Figure 7 shows these patterns qualitatively. Figure 7 also 
shows that $n$ affects the light curve steepening: with the increase of 
$n$, the time that the ejecta enters the Newtonian stage comes earlier, 
so that the light curve break becomes less and less striking. 
The effect of dense media on afterglows from isotropic fireballs 
has been discussed by Dai \& Lu (1999a, b).

Figure 8 illustrates the effect of $p$ on the light curves. From 
equation~(\ref{tm14}) we know that $t_{\rm m}$ is related to $p$. Our 
numerical results reflect the relation correctly. Generally speaking, 
although $p$ affects the slope of the light curve notably, it has 
minor effect on the light curve steepening. This again is very different 
from that of an isotropic fireball, whose timing index $\alpha$ varies 
from $(3p-3)/4$ in the relativistic phase to $(15p-21)/10$ in the 
Newtonian phase (Wijers, Rees \& M\'{e}sz\'{a}ros 1997; 
Dai \& Lu 1999a, b), i.e., the increment $\Delta \alpha$ 
strongly depends on $p$: $\Delta \alpha = (27-15p)/20$. 

Light curves in Figure 9 are drawn under different assumptions. The 
solid line corresponds to a ``standard'', ``ideal'' jet (adiabatic), 
the dashed line corresponds to a ``realistic'' jet (whose radiative 
efficiency $\epsilon$ evolves according to equation~(\ref{eps6})), 
and the dash-dotted line corresponds to a jet without lateral 
expansion (i.e., $c_{\rm s} \equiv 0$). It is generally believed 
that the lateral expansion effect tends to make the light curve 
steepening more obvious. However, Figure 9 has just shown the 
opposite tendency: the steepening of the dash-dotted line is obviously 
more striking than that of the solid line. In fact the numerical 
simulation by Moderski, Sikora \& Bulik (1999) also shows this 
tendency. From this we conclude that it is the jet edge effect 
that leads to the additional light curve steepening (with respect 
to an isotropic fireball) at the relativistic-Newtonian transition 
point. The lateral expansion effect tends to reduce the steepening. 

Figure 10 shows the evolution of the observed spectra for the 
``standard'' jet. At early times, the spectrum can be divided into 
three segments. But at time of $t \sim 10^6$ s, a fourth segment 
emerges. It may come from the edge effect in Newtonian phase, 
and it is just this segment that leads to the light curve steepening.

\begin{figure} \centering
\epsfig{file=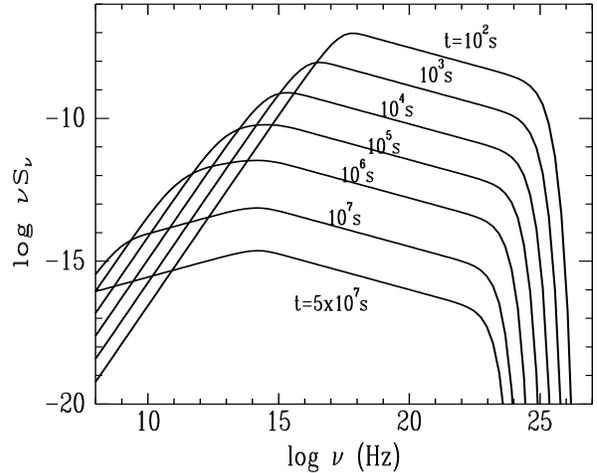, angle=-90, height=7cm, width=8.5cm,
bbllx=130pt, bblly=70pt, bburx=570pt, bbury=640pt}
\caption{The spectral evolution of a jet with ``standard'' parameters.
$\nu S_{\nu}$ is in units of ergs cm$^{-2}$ s$^{-1}$. The curves from
upper-right to lower-left correspond to $t = 10^2, 10^3, 10^4, 10^5,
10^6, 10^7$, and $5 \times 10^7$ s respectively.}
\end{figure}

\section{Conclusion}

We have studied the evolution of jetted GRB remnants numerically, 
following the convenient model developed by Huang et al. (1999c). 
In typical cases, the remnant enters the mildly relativistic 
phase at time $t \sim 10^5$ --- $10^6$ s (i.e., 1 --- 10 d), and it 
will become non-relativistic at time $t \geq 10^{6.5}$ s (37 d). In 
analytic approximation, it is usually assumed that the expansion 
is highly relativistic ($\gamma \gg 1$) all the time. Now we should 
note that these approximations are appropriate only for early afterglows 
(Huang et al. 1998a, b, 1999a, b, c). Due to lateral expansion, the 
half opening angle $\theta$ will increase with time. But the increment 
of $\theta$ is very small at the ultra-relativistic stage. After 
entering the Newtonian stage, the increase of $\theta$ is very quick. 

It has been predicted that due to the edge effect 
and the lateral expansion effect, 
the light curve of afterglows from jetted GRB remnants should show 
obvious steepening during the relativistic phase, with the break 
point determined by $\gamma \sim 1/\theta$. However, our numerical 
results show clearly that no steepening could be observed during 
{\em the relativistic stage itself} (also see Huang et al. 1999c). 
But in many cases, a striking break does appear in the light curve 
at later stages, which is obviously due to the relativistic-Newtonian 
transition. Typically the power law timing index $\alpha$ varies 
from $\sim 1.2$ to $\geq 2.1$. 
 
The light curve break due to the relativistic-Newtonian transition 
is affected by many parameters, such as $\xi_{\rm e}, \xi_{\rm B}^2, 
\theta_0$ and also $n$. Increase of any of them to a large enough 
value will make the break disappear. We have also demonstrated that 
it is the jet edge effect (Kulkarni et al. 1999; M\'{e}ez\'{a}ros 
\& Rees 1999) that leads to the light curve break. The lateral 
expansion effect just tends to reduce the steepening (see Figure 9), 
this may be completely unexpected to most researchers.

Although whether the break appears or not depends on parameters, 
afterglows from jetted GRB remnants are uniformly characterized by 
a quick decay during the non-relativistic phase, with $\alpha 
\geq 2.1$. This is quite different from isotropic fireballs, whose 
light curves steepen only slightly after entering the non-relativistic 
phase. It can be regarded as {\em a fundamental characteristic of jets}
and may be used to judge the degree of beaming. 

The effect of $p$ on the light curve break is also very interesting.
For an isotropic fireball, if $p$ is as small as 2.1, the light curve 
will steepen only slightly after entering the non-relativistic phase. 
But for a jet, if other parameters are properly assumed, we still can 
observe a steep break. In practical observations, if $p$ can be 
determined to be near 2 from spectral information, and if we observed 
a steep break in the light curve, then possibility is large that we 
were observing afterglows from a jet.

\section{Discussion}

We have shown that the afterglow from jetted GRB remnant is characterized 
by a steep light curve (with $\alpha \geq 2$) at late stages. However, 
beaming is not the only factor that leads to such a steep light curve. 
If the GRB progenitor is a massive star, a stellar wind environment will 
be created, where the density scales as $n \propto R^{-2}$. Then the 
afterglow light curve will also be quite steep (Chevalier \& Li 1999; 
Halpern et al. 1999; Frail et al. 1999; Dai \& Lu 1998a). 
This makes the problem even more complicated. 

In practical afterglow observations, we may encounter four kinds of 
optical light curves: (a) a single flat line, with $\alpha \sim 1.1$; 
(b) a single steep line, with $\alpha \geq 2$; (c) a slightly broken 
curve, with $\alpha$ evolves from $\sim 1.1$ to $\sim 1.5$; (d) a 
steeply broken curve, with $\alpha$ evolves from $\sim 1.1$ to $\geq 2.1$. 
let us discuss their meanings one by one:
\begin{description}
\item[{\bf (a)}] {\bf A single flat line, with $\alpha \sim 1.1$.} In this 
   case, we can safely say that the ejecta is not highly beamed. 
   GRB 970508, 971214, 980329 and 980703 may be good examples. 
\item[{\bf (b)}] {\bf A single steep line, with $\alpha \geq 2$.} This may be 
   due to either a highly collimated jet or an isotropic fireball in a 
   wind environment. We can not distinguish them solely by optical light 
   curve features. Frail et al. (1999) suggested that they can be 
   distinguished by radio afterglows. GRB 980326, 980519 and probably 
   GRB 970228 belong to this case.  
\item[{\bf (c)}] {\bf A slightly broken curve, with $\alpha$ evolves from 
   $\sim 1.1$ to $\sim 1.5$.} In this case the ejecta should be isotropic, 
   and the break is likely due to the relativistic-Newtonian transition of 
   the remnant. 
\item[{\bf (d)}] {\bf A steeply broken curve, with $\alpha$ evolves from 
   $\sim 1.1$ to $\geq 2.1$.} In this case, the ejecta is probably 
   highly collimated. The break is due to the relativistic-Newtonian 
   transition of the jet. 
\end{description}
Showing obvious breaks in the optical light curves, GRB 990123 and 990510 
are the most hopeful candidates for beaming. However their late time 
timing index are still not accurately determined, we cannot definitely say 
that they belong to case (c) or (d). To determine the late time 
index accurately, we need a well defined optical light curve with 
$t \geq 30$ d. This is a difficult task. 

>From above discussions, we see that till now we have observed at least 
three kinds of optical light curves. They belong to case (a), case (b) 
and case (c) or (d) respectively. This means that there are at least two 
kinds of GRB afterglows: one corresponds to an isotropic fireball in 
a uniform medium (i.e., case (a) ), the other corresponds to an isotropic 
fireball in a wind environment or a jet. These kinds of information can 
provide important clues to our understanding of GRBs. For example, at 
least we know that $\gamma$-ray emission in some GRBs is isotropic, 
then the energy crisis is really a problem: GRB 971214 and 980703 
indicate isotropic energies of $\sim 0.17 M_{\odot} c^2$ and  
$\sim 0.06 M_{\odot} c^2$ respectively. 

Systematic deep optical surveys may provide another way for determining 
the degree of beaming. If GRBs are highly collimated, these surveys may
reveal many faint decaying optical sources. They are afterglows from 
jetted GRBs whose $\gamma$-ray emission deviates the line of sight 
slightly. 

\section*{Acknowledgments}
This work was partly supported by the National Natural Science 
Foundation of China, grants 19773007 and 19825109, and the National Climbing 
Project on Fundamental Researches.

{}

\end{document}